\documentclass[structabstract]{aa}  
\usepackage{graphicx}
\usepackage{txfonts}
\usepackage{longtable}
\begin{document}
   \title{Spectroscopic orbits and variations of RS Oph}

%\subtitle{ }

   \author{E. Brandi
          \inst{1,}\inst{2,}\inst{3}
\and 
         C. Quiroga\inst{1,}\inst{2} %\fnmsep\thanks{Just to show the usage of the 
                                       %elements in the author field}
\and 
         J. Miko{\l}ajewska\inst{4}
\and           
O.E. Ferrer\inst{1,}\inst{5}
          \and
          L.G. Garc\'ia\inst{1,}\inst{2}
          }

   \institute{Facultad de Ciencias Astron\'omicas y Geof\'isicas, Universidad Nacional de La 
                 Plata (UNLP), Argentine \\
              \email{ebrandi@fcaglp.unlp.edu.ar}
  \and  Instituto de Astrof\'isica de La Plata (CCT La Plata-CONICET-UNLP), Argentine
% \email{}
\and  Comisi\'on de Investigaciones Cient\'{\i}ficas de la Provincia de 
            Buenos Aires (CIC), Argentine 
%\email{}    
  \and
    Copernicus Astronomical Center, Warsaw, Poland \\
             \email{mikolaj@camk.edu.pl}
\and  Consejo Nacional de Investigaciones Cient\'{\i}ficas y T\'ecnicas (CONICET), 
Argentine
             %\thanks{The university of heaven temporarily does not accept e-mails}
             }           

   \date{Received ; accepted }

% \abstract{}{}{}{}{} 
% 5 {} token are mandatory
 
  \abstract
 {}  
   {The aims of our study are to improve the orbital elements of the giant, and to derive 
the spectroscopic orbit for the white dwarf companion.  Spectral variations related to the 
2006 outburst are also studied.}
   { We performed an analysis of about seventy optical and near
infrared spectra of RS Oph that were acquired between 1998 and June 2008.
The spectroscopic orbits have been obtained by measuring the radial
velocities of the cool component absorption lines and the broad
H$\alpha$ emission wings, which seem to be associated with the hot component. 
A set of cF-type absorption lines were also analyzed for a possible connection with the hot component motion.
}
   {A new period of 453.6 days, and a mass ratio, $q=M_{\rm g}/M_{\rm h}=0.59 \pm 0.05$, 
were determined. Assuming a massive white dwarf as the hot component 
($M_{\rm h}= 1.2 - 1.4\rm M_{\odot}$) the red giant mass is 
$M_{\rm g}= 0.68 - 0.80 \rm M_{\odot}$ 
and the orbit inclination, $ i = 49^{\circ} - 52^{\circ}$.
The cF-type lines are not associated with either 
binary component, and are most likely formed in the material streaming towards the hot component.
We also confirm the presence of the Li\,I doublet in RS Oph and its radial velocities 
fit very well the M-giant radial velocity curve. Regardless of the mechanism  involved 
to produce lithium, its origin is most likely from within the cool giant
rather than material captured by the giant at the time of the nova
explosion.

The quiescent spectra reveal a correlation of the H\,I and He\,I emission line fluxes with the 
monochromatic magnitudes at 
4800 \AA\ indicating that the hot component activity is responsible for those flux variations. 
We also discuss  the spectral characteristics around 54--55 and 240 days after the 2006 outburst. 
 In April 2006  most of the emission lines present a broad pedestal with a 
strong and narrow component at about -20 $\rm km\,s^{-1}$ and two other extended emission components at 
-200 and +150 $\rm km\,s^{-1}$. These components could originate in a bipolar gas outflow supporting 
the model of a bipolar shock-heated shell expanding through the cool component wind 
perpendicularly to the binary orbital plane.
Our observations also indicate that the cF absorption system was disrupted during the outburst, and 
restored about 240 days after the outburst, which is consistent with the resumption of accretion.}
{} 


 \maketitle
%
%________________________________________________________________

\section{Introduction}
 RS Oph is a recurrent symbiotic nova, in which a white dwarf near the
Chandrasekhar limit  orbits inside the outer wind of a red giant. 
The system has had numerous recorded 
outbursts (1898, 1933, 1958, 1967, 1985). The most recent outburst began on 
2006 February 12 and multifrequency observations confirmed the current model 
for the outburst in which the massive white dwarf accretes material from the 
red giant until a thermonuclear runaway ensues and high velocity gas is 
ejected from the white dwarf. Because the white dwarf is orbiting inside
the outer layers/wind of the red giant, a strong shock structure is established
around the system.\\
 For RS Oph, Garcia (1986) made the first orbital period estimate of about 
230 days, based on eight radial velocity measurements. Two sets of lines were 
considered: 
M star absorption features and absorption cores of \ion{Fe}{ii} emission lines due to a 
shell. He could not, however, derive the mass ratio.
Oppenheimer \& Mattei (1993) analysed intervals between outbursts in the AAVSO 
visual light curve and found  the strongest periods ranged between 892 and 2283 days.
Dobrzycka \& Kenyon (1994) (hereafter DK94) continued monitoring the radial velocities 
of RS Oph during quiescence. They also separated their echelle spectra into sets of M-type and A-type absorption lines, both of which resulted in a period of 460 days. 
The A-type 
radial velocity curve was shifted by 0.37 relative to the M giant solution and
it was not associated with the hot component motion. They concluded that their 
results are more reasonable if they only considered the circular orbit of the 
cool giant. DK94 searched for periodic variations in photometric data of RS Oph 
collected during 1971--1983 and after the 1985 outburst. For both time intervals
they found the light curve to be the superposition of a longer period variation
(P=2178 days) and a shorter variation (P=508 days). The longer period was in agreement 
with P=2283 days given by Oppenheimer \& Mattei (1993) but it did not appear in their 
radial velocity analysis so it could not be associated with the orbital motion.
Fifteen infrared radial velocities were used by Fekel et al.(2000) (hereafter FJHS00)
combined with the 47 optical velocities of DK94 in order to improve the orbital elements 
of the giant. FJHS00 fitted a new period of 455.72 days for a circular solution and 
eccentric orbits were rejected applying the test of Lucy \& Sweeney (1971).

In this study, we have collected spectroscopic data of RS Oph from September 1998 until October 2006 and two 
Feros spectra taken on May and June 2008 were also included in our analysis.
Our observation closest to the eruption was obtained in April 2006. \\
We describe our observations in Sec. 2; in Sec. 3 we determine double-line spectroscopic 
orbits for RS Oph; in Sec. 4 spectral characteristics observed during quiescence 
are presented;
in Sec. 5 we describe spectral changes during the active phase at  54--55 and 240 days after 
eruption and finally a brief discussion of our results and conclusions are given in Sec. 6.

%__________________________________________________________________

\section{Observations}
\subsection{Spectroscopy}

Spectroscopic observations of RS Oph were performed with the 2.15 m ''Jorge Sahade'' 
telescope of CASLEO\footnote{Complejo Astron\'omico El Leoncito operated under agreement 
between the Consejo Nacional de Investigaciones Cient\'ificas y T\'ecnicas de la Rep\'ublica
Argentina and the National Universities of La Plata, C\'ordoba and San Juan} 
(San Juan, Argentina), during 1998--2006. 65 spectra with  
resolution 12000--15000 were obtained with a REOSC echelle spectrograph using a Tek CCD 
1024$\times$1024 pixels.
The CCD data were reduced with IRAF\footnote{IRAF is distributed by the National Optical
Astronomy Observatories, which is operated by the association of Universities
for Research in Astronomy, INC. under contract to the National Science
Foundation} packages, CCDRED and ECHELLE and all the spectra were 
measured using the SPLOT task within IRAF.

To obtain the flux calibration, standard stars from Hamuy et al. (1992) and Hamuy et al.(1994)
were observed each night. A comparison of the spectra of the standards suggests that the 
flux calibration errors are about 20\% in the central part of each echelle order.

Two high resolution spectra were acquired in May and June 2008 with 
the Feros spectrograph mounted at the 2.2m telescope at ESO, La Silla (Chile).
An EEV CCD-44  with 2048$\times$4096 pixels was used as detector with a pixel size of 
15$\times$15 $\mu$m and resolving power of $R=60000$ .

\section{Spectroscopic orbits}

Spectroscopic orbits of RS Oph were calculated using  high resolution spectra
collected at CASLEO during the period 1998--2006.

%______________________________________________FIG.1 
   \begin{figure*}
   \centering
    \includegraphics[width=11cm,angle=90]{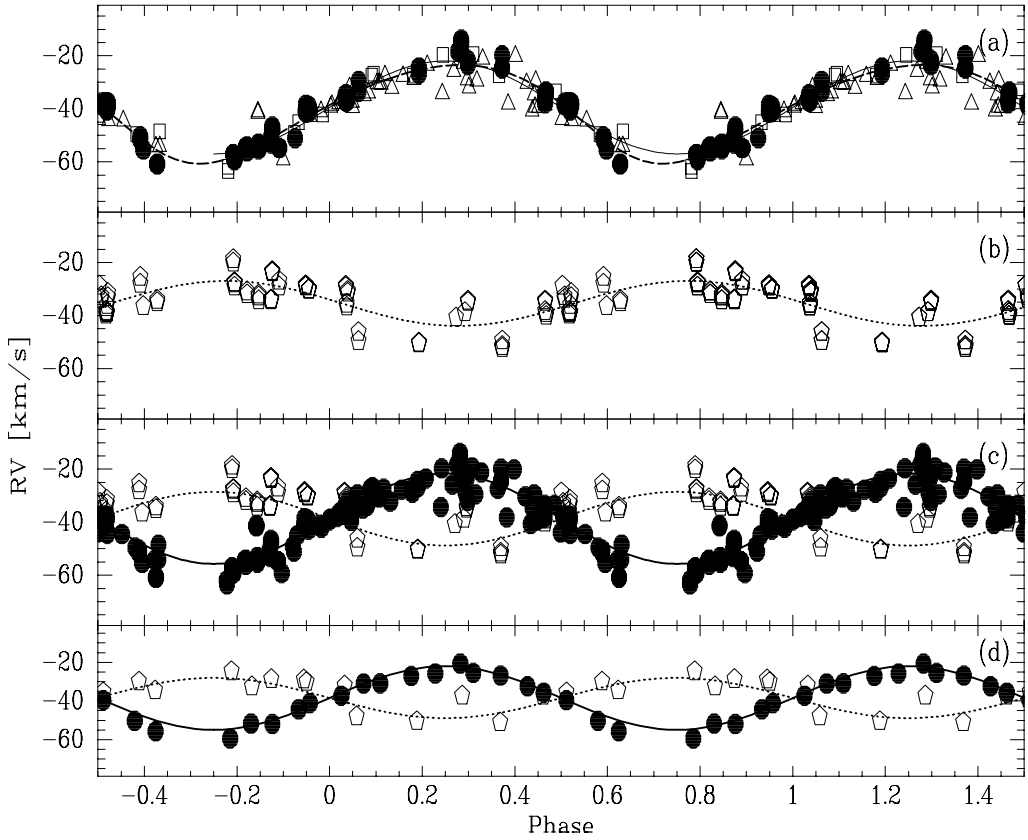}
   \caption {Radial velocity curves and orbital solutions
for RS Oph. The data are phased with the period of 453.6 days and 
$T_{\it conj} = 2445043.54\pm5$ 
(the time of the inferior conjunction of the M giant).
(a) M-giant. Filled circles represent our data and open triangles and
squares, the DK94's and  FJHS00's data, respectively. Solid line gives 
the best circular fit and dashed line gives the eccentric orbit (e=0.14). 
(b) The H$\alpha$ wings (hereafter open symbols) and the best circulat solution (dotted line).
(c) Combined circular solution for the M-giant (filled circles) and the H$\alpha$ 
     wings. 
(d) The same with binned data.
}
  \label{fig.1}
    \end{figure*}

%_____________________________________________________

To obtain the radial velocities of the red giant we have measured
M-type absorptions in the region  $\lambda\sim$6000--8000 \AA, corresponding to 
\ion{Fe}{i}, \ion{Ti}{i}, \ion{Ni}{i}, \ion{Si}{i}, \ion{O}{i}, \ion{Zr}{i}, 
\ion{Co}{i}, \ion{V}{i}, \ion{Mg}{i} and \ion{Gd}{ii}. 
 We have also identified and measured absorption lines of singly ionized elements in the blue region of our spectra ($\lambda\sim$4000--5800 \AA), which resemble spectra of A--F supergiants. This set of absorption lines, called cF-absorptions, are believed to be linked to the hot companion (e.g. Miko{\l}ajewska \& Kenyon 1992; Quiroga et al. 2002; Brandi et al. 2005). These are basically the same lines that are called A-type absorption lines
by DK94. As the \ion{Fe}{ii} lines show very variable and complex profiles with emission components, only the stronger \ion{Ti}{ii} absorption lines were considered in our orbital solutions.\\
In both cases individual radial velocities were 
obtained by a Gaussian fit of the line profiles. A mean value was calculated for each 
spectrum and the resulting heliocentric velocities together with their standard errors 
and with the number of measured lines are given in Table~\ref{rv}.  The published radial velocities of the M-giant were also included in the table.

\addtocounter{table}{1}

Table~\ref{orbital-sol} lists the resulting orbital solutions for the M-giant using 
only our radial velocities as well as our radial velocities 
combined with the DK94's and  FJHS00's data and considering in this case weighted 
values of radial velocities.  We have applied a weight 1 to the FJHS00 data and
weights 0.33, 0.67 and 1 when the errors are larger than 4, between 2 and 3 and less than 2,  respectively, for both the DK94 and our data. But we noted that the weighted solutions
show no important differences when a weight 1 is applied to all the data.\\
A new orbital period of 453.6$\pm$0.4 days  was determined. An elliptical orbit 
({\it e}=0.14$\pm$0.03) fits the measured velocities slightly better than the 
circular one (see Fig.~\ref{fig.1}a). \par

\begin{figure} 
\resizebox{\hsize}{!}{\includegraphics[angle=90]{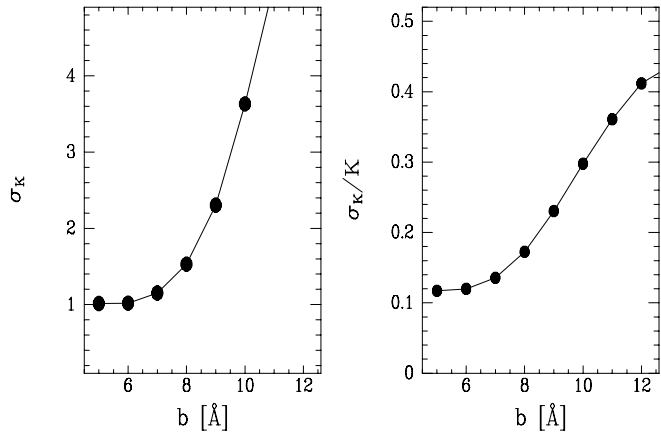}}
\caption{The standard error of the orbital semi-amplitude $\sigma_K$ 
and the $\sigma_K/K$ ratio as a function of the parameter $b$. The best 
estimation of $K$ is determined by the value of $b$ where $\sigma_K$ begins 
to sharply increase.}
\label{fig.2} 
\end{figure}

 We have also determined the radial velocities from the broad emission wings of 
H$\alpha$ which should reflect the motion of the hot component if they were formed  in 
the inner region of the accretion disk or in an extended envelope near the hot component. 
 For this we have used the method outlined by Schneider \& Young (1980) 
(for more details see Quiroga et al. 2002).  This method consists of convolving the data with two identical Gaussian bandpasses whose centers have a separation of $2b$ with standard deviation $\sigma$. We have experimented with a range of values for $b$ and $\sigma$ fixed at 3\AA, for which we have obtained the circular orbit solutions. Fig.~\ref{fig.2} shows the dependence of the standard error $\sigma_K$ of the orbital semi-amplitude $K$, and the $\sigma_K/K$ ratio, as a function of $b$. It is noticeable that $\sigma_K$ and $\sigma_K/K$ increases sharply for $b$ larger than $\sim$ 7\AA. We can attribute these large standard errors to the velocity measurements dominated by the noise in the continuuum rather than by the extreme high velocity wings of the line profile. We have therefore adopted $b=7\AA$ as the best value. The heliocentric radial velocities of the broad emission wings of H$\alpha$
are included in Table~\ref{rv}.\\
The broad emission line wings of H$\alpha$ show a mean velocity similar to the red giant 
systemic velocity (see Fig.~\ref{fig.1}b). They are clearly in antiphase with the 
M-giant curve, which suggests that they are really formed in a region very near to the hot component.\par 
%______________________________________________________FIG.2

%_______________________________________________________________

Finally we have calculated the combined orbital solutions from the M-giant absorptions 
and from the H$\alpha$ broad emission wings (see Table~\ref{orbital-sol}).  From a comparison of the variances of the velocities of the two individual solutions, 
weights of 0.64 and 1 were given to the H$\alpha$ wings and the M-giant velocities, respectively. 
Leaving the period as a free parameter and taking a circular orbit, the best solution 
corresponds to a similar period of 454.1$\pm$0.41 days 
and adopting the period  453.6 days, the best combined solution leads to a lower 
eccentricity, $e=0.04\pm0.03$.
Moreover,  we have considered binned data of radial velocities taking phase intervals 
of 0.05, and as it is shown in Table~\ref{orbital-sol}, very similar 
results were obtained. 
These solutions show a clear antiphase variations of both sets of radial velocities 
and the systemic velocity agrees very well with that of the cool component orbital 
solutions. 
The combined unbinned and binned solutions are shown in Fig.~\ref{fig.1}c and d, 
respectively. \par
 
Using the combined circular solutions for unbinned data, we have estimated the  mass ratio 
$q=M_{\rm g}/M_{\rm h}=0.59\pm 0.05$,  the component masses 
$M_{\rm g} \sin^3 i = 0.35\, \rm M_{\odot}$ and $M_{\rm h} \sin^3 i = 
0.59\, \rm M_{\odot}$, and the binary separation $a \sin i = 240\, \rm R_{\odot}$.
Assuming that the hot component is a massive white dwarf close to the Chandrasekhar
limit ($M_{\rm h}= 1.2$--$1.4\,\rm M_{\odot}$), we estimate the red giant mass, 
$M_{\rm g}= 0.68$--$0.80\,\rm M_{\odot}$ and the orbit inclination, 
$ i = 49^{\circ}$--$52^{\circ}$, which is consistent with the absence of eclipses
in both, the optical light curve and the \ion{H}{i} and \ion{He}{i} emission line 
fluxes in our spectra.\par

A lower limit for the mass ratio can be also derived from  the ratio of the rotational 
velocity to the orbital semiamplitude of the red giant and assuming that the giant is 
synchronized. 
Adopting $v \sin i =11.7 \pm 1.5\, \rm km\,s^{-1}$ (Zamanov et al. 2007) and $K_{\rm g}=17.1 \pm 0.6\, \rm km\,s^{-1}$ 
(Table~\ref{orbital-sol}) results in $q_{\rm min}=0.68 \pm 0.1$.  This lower limit 
is surprisingly close to the mass ratio derived from the radial velocity curves, and 
indicates that either the giant is filling its Roche lobe (RL) -- the actual $q$ should then be 
equal to $q_{\rm min}$ -- or that its measured rotational 
velocity is faster than the synchronized value. The first possibility is favored by 
theoretically predicted synchronization and circularization time-scales for convective 
stars (Zahn 1977). 
It is also easier to ensure the high mass tranfer and accretion rate required by the 
activity and short nova outburst recurrence time of RS Oph. On the other hand, 
the presence of a RL-filling giant implies the distance to RS Oph is a factor of 2 
larger than generally accepted based on observational evidence (Barry et al. 2008).

\begin{figure}
   \centering
    \includegraphics[width=8cm,angle=0]{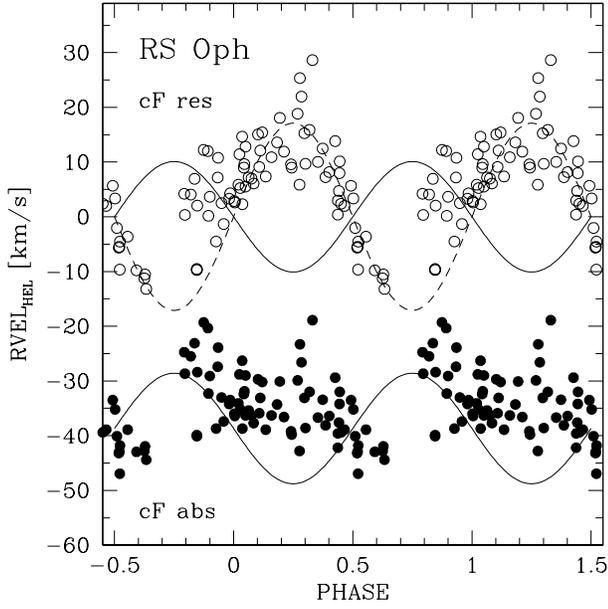}
   \caption {Phased radial velocities of the cF-absorption lines (dots) and the best circular solution 
(solid line) for the hot component based on the H$\alpha$ wings ({\it bottom}).
The departures of the cF-abs radial velocities from the circular orbit of the hot 
component  are due to a significant contribution from the gaseous stream between
both stars. The residuals ($O-C$) are represented by empty circles whereas the solid and dashed lines mark the 
best circular solution for the hot and cool component, respectively ({\it top}).
}
  \label{fig.3}
    \end{figure}

The cF absorption lines do not trace clearly the orbit of the hot component. Any orbital 
solution leads to a significant eccentricity (see Table~\ref{orbital-sol}) and the radial 
velocity curve is shifted by $\sim-0.26\,P$ (0.74\,$P$) relative to the M giant solution 
instead of 0.5\,$P$ (see Fig.~\ref{fig.3}). We think, in agreement with DK94, that in 
RS Oph the cF absorption lines are not associated with either binary component, and are most 
likely formed in the material streaming towards the hot component. We note here that 
similar complications 
with the blue absorption system occurred for a few other active symbiotic systems. 
Miko{\l}ajewska \& Kenyon (1996) failed to derive any radial velocity curve and orbital 
solution for the blue absorption system in Z And. In CI Cyg the radial velocities of the 
F-type absorption system suggest that their formation region is the material streaming 
from the giant 
near the hot component (Miko{\l}ajewska \& Miko{\l}ajewski 1988), whereas a possible 
correlation between AR Pav's activity and the departures of the cF absorption velocities 
from the circular orbit suggests that this absorption system may be also affected by 
material streaming towards the hot component presumably in a region where the stream 
encounters an accretion disk or an extended envelope around the hot component 
(Quiroga et al. 2002). We can see in Fig.~\ref{fig.3} the departures 
of the cF-abs radial velocities from the circular orbit of the hot component due to a 
significant contribution from the stream. 

 In Fig.~\ref{fig.1} and Fig.~\ref{fig.3} and hereafter we adopt the period of 453.6 
days and the origin of the phases corresponds to the inferior conjunction of the 
M-giant:

      \[  T_{\it conj}= JD\, 2445043.54 \pm5 + 453.6 \pm0.4\, E \]

%__________________________________________________FIG.3

%________________________________________________Table 2 - One column table

\begin{table*}
\caption{Orbital solutions of RS Oph}
\smallskip
\label{orbital-sol}
\begin{center}
\begin{tabular}{lcccccccll} 
\hline
\noalign{\smallskip}
Component & P  &  $K$              & $\gamma_0$ &     $e$     & $\omega$& $T_0^{(1)}$& $\Delta$$\phi^{(2)}$ & $a\sin i$ & $f(M)$ \\
        &[days]& [$\rm km\,s^{-1}$] & [$\rm km\,s^{-1}$] &             & [deg]  &  [JD\,24...]     &           &[AU]    &[M$_\odot$]\\
%\noalign{\smallskip}
\hline
M-giant   &                     &               &    &            &          &     &      &     \\
Our data  & 452.8$\pm$1.5 & 19.5$\pm$0.7 & -38.0$\pm$0.4 & 0$^{(3)}$ &            & 51055.50 &     & 0.81 & 0.35 \cr  
Our data  & 453.2$\pm$1.2 & 20.0$\pm$0.7 & -38.7$\pm$0.4 & 0.10$\pm$0.03& 120$\pm$16 & 51205.95 &     & 0.83 & 0.37 \cr         
DK+F+ours & 453.6$\pm$0.4 & 17.5$\pm$0.6 & -39.5$\pm$0.4 & 0$^{(3)}$&            & 45156.94 & 0.00& 0.73 & 0.25    \cr
DK+F+ours & 453.6$^{(3)}$& 18.6$\pm$0.6 & -40.2$\pm$0.4 & 0.14$\pm$0.03& 135$\pm$11  & 44873.23 &     & 0.77 & 0.29 \cr 
\hline
 cF-abs &             &              &               &              &            &          &     &      &   \cr
Our data$^{(5)}$ & 453.6$^{(3)}$ & 8.0$\pm$1.2  & -32.5$\pm$0.9 & 0$^{(3)}$   &            & 50493.51 & -0.24&0.33 & 0.024  \cr
                 & 453.6$^{(3)}$ & 10.2$\pm$1.8 & -31.8$\pm$0.8 & 0.52$\pm$0.14& 200$\pm$13 & 50750.54 &      &0.36 & 0.031  \cr 
DK+ours          & 453.6$^{(3)}$ & 3.2$\pm$0.9  & -34.7$\pm$0.7 & 0$^{(3)}$  &            & 45041.11 & -0.26&0.13 & 0.001  \cr
                 & 453.6$^{(3)}$ & 8.3$\pm$2.8  & -34.6$\pm$0.7 & 0.82$\pm$0.07& 344$\pm$16 & 46799.95 &      &0.20 & 0.005 \cr
\hline
Emission wings  & &            &               &              &             
&          & &        &    \cr
H$\alpha$ wings       & 456.0$\pm$6.1&  8.6$\pm$1.2 & -35.4$\pm$0.70&0$^{(3)}$ &     & 50376.11 &  &  &  \cr
H$\alpha$ wings       & 453.6$^{(3)}$&  8.5$\pm$1.2 & -35.4$\pm$0.70&0$^{(3)}$ &     & 50385.80 & 0.53 &0.42 & 0.029 \cr
H$\alpha$ wings$^{(4)}$& 453.6$^{(3)}$& 10.0$\pm$2.3 &-36.6$\pm$1.5 & 0$^{(3)}$ &    &          &     &0.35 & 0.047 \cr
\hline
\end{tabular}

Combined orbital solutions ($P=453.6\pm0.4$ days)\\
%\smallskip
%\tabcolsep2pt
\begin{tabular}{lccccccc} 
\hline\hline
\noalign{\smallskip}
Component  & $K$ & $\gamma_0$  & $e$   & $\omega$& $T_0^{(1)}$& $a\sin i$  &$M\sin^3$i \\
           & [$\rm km\,s^{-1}$]& [$\rm km\,s^{-1}$] & & [deg] & [JD24...] & [AU]  &  [M$_\odot$] \\ 
\noalign{\smallskip}
\hline
\noalign{\smallskip}
DK+F+ours       &  17.1$\pm$0.6 & -38.7$\pm$0.4 & 0$^{(3)}$ &  & 45612.05& 0.71 & 0.35 \\
H$\alpha$ wings &  10.1$\pm$1.2 &               &           &  &         & 0.42 & 0.59 \\
\hline
DK+F+ours       &  17.2$\pm$0.7 & -38.7$\pm$0.4 & 0.04$\pm$0.03 & 87$\pm$44& 45722.37 &0.72 & 0.35\\
H$\alpha$ wings &  10.2$\pm$1.2 &               &               &          &          &0.42 & 0.59\\
\hline
DK+F+ours$^{(4)}$       & 16.4$\pm$1.5 & -38.4$\pm$0.8 & 0$^{(3)}$ &   & 0.00 & 0.68 & 0.35 \\
H$\alpha$ wings$^{(4)}$ & 10.4$\pm$1.8 &               &           &   &      & 0.43 & 0.55 \\
\noalign{\smallskip}
\hline
\end{tabular} 

\end{center}
Notes: (1) $T_0$ is the time of maximum velocity (circular orbits) 
or the time of periastron passage (elliptical orbits)\\ 
(2) $\Delta\phi$= ($T_0$ - $T_{0, \rm giant}$)/P \\
(3) assumed \\
(4)  binned data \\ 
(5) solutions using only the measurements of the \ion{Ti}{ii} absorption lines.

\end{table*}
%_______________________________________________________FIG4

 \begin{figure*}
   \centering
   \includegraphics[width=9cm,angle=90]{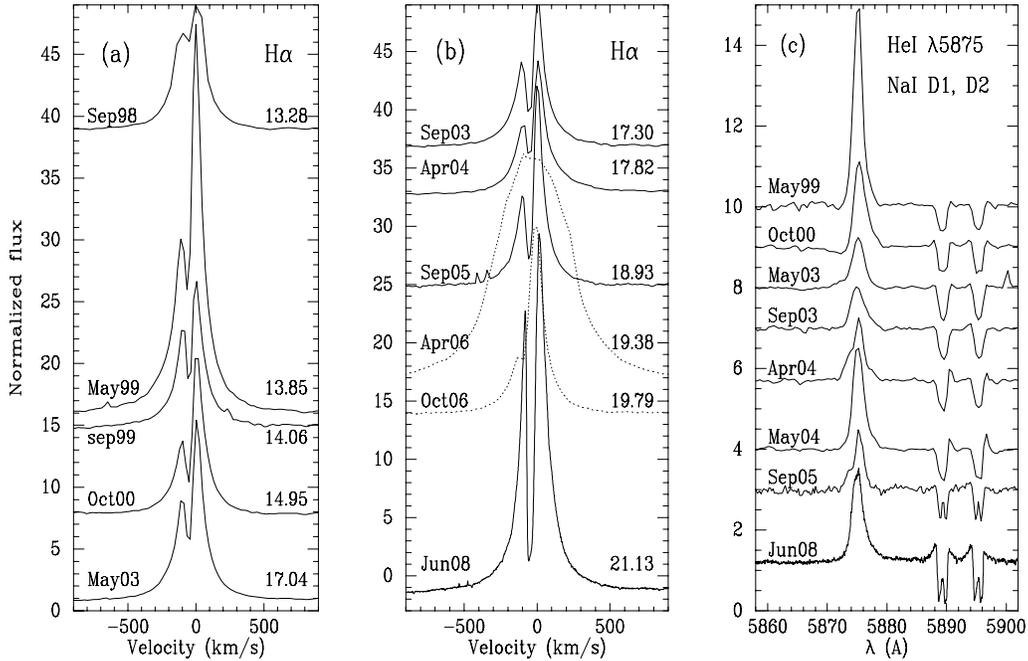}
      \caption{({\it panels (a) and (b)}): Profiles of H$\alpha$. At 
quiescence the broad emission is cut by a central absorption which disappears 
during the outburst (dotted profile). 
({\it panel (c)}): Profile variations of \ion{He}{i}\,$\lambda 5875$ and \ion{Na}{i} 
doublet during the quiescent period. 
Blueshifted broad components in \ion{He}{i} are present on April 2004 and September 2005.
 The \ion{Na}{i} doublet shows weak emission components and double structure in the absorptions.           
}
         \label{fig.4}
   \end{figure*}

%_________________________________________________________Table 3

\begin{table}
\caption{Mean values of radial velocities and equivalent widths of the central 
absorptions in H$\alpha$ and H$\beta$}
\tabcolsep1pt
\label{rv}
\begin{center}
\begin{tabular}{lccccc} 
\hline\hline
JD & Phases  &\multicolumn{2}{c}{H$\alpha$}& \multicolumn{2}{c}{H$\beta$} \\
2450000+& (days) & RV($\rm km\,s^{-1}$)& eqw  & RV($\rm km\,s^{-1}$)& eqw \\ 
\hline  
 1069.6 & 13.285 & -57.7& 0.46 & -25.6 &  0.43 \\
 1324.7 & 13.847 & -56.6& 0.59 & -47.9 &  0.57 \\
 1422.6 & 14.063 & -50.2& 0.53 & -43.0 &  0.72\\
 1629.9 & 14.518 & -48.8& 0.71 & -45.0 &  0.60 \\
 1754.7 & 14.795 & -52.3& 0.70 & -46.9 &  0.58\\
 1790.6 & 14.874 & -52.4& 0.73 & -45.1 &  0.61\\
 1791.6 & 14.877 & -50.0& 0.67 &       &      \\
 1824.5 & 14.949 & -50.7& 0.74 & -47.1 &  0.66\\
 1827.5 & 14.956 & -52.7& 0.72 &       &      \\
 2016.9 & 15.373 & -48.4& 0.75 & -45.2 &  0.65 \\
 2059.9 & 15.468 & -43.8& 0.59 & -47.8 &  0.60 \\
 2132.6 & 15.628 & -48.1& 0.62 & -54.7 &  0.64 \\
 2388.9 & 16.193 & -42.9& 0.86 & -43.4 &      \\
 2531.6 & 16.508 & -38.8& 0.77 &       &        \\
 2770.9 & 17.036 & -48.4& 0.63 & -48.5 &  0.76\\
 2771.8 & 17.038 & -44.4& 0.76 & -41.1 &  0.86\\
 2890.5 & 17.299 & -50.1& 0.86 &      &      \\
 3126.8 & 17.820 & -48.4& 0.68 & -40.8 &  0.56\\
 3127.8 & 17.822 & -47.3& 0.83 &       &     \\
 3476.9 & 18.592 & -43.8& 0.93 & -42.9 &  0.75 \\
 3628.6 & 18.926 & -51.1& 0.92 & -48.2 &  0.79\\
 4600.8 & 21.070 & -60.0& 0.94 & -54.5 &  0.88\\
 4627.8 & 21.129 & -43.6& 1.07 & -35.2 &  0.95\\
\hline
\end{tabular}
\end{center}
\end{table}

\section {RS Oph at quiescence} 

Fluctuations in the fluxes of the emission lines and the
continuum are observed during the period of quiescence 
covered by our observations. Several authors have previously reported such behaviour 
of RS Oph, outside the eruptive episodes (see Anupama \& Miko{\l}ajewska, 1999 
and refences therein).

\begin{figure*}
   \centering
    \includegraphics[width=5cm,angle=90]{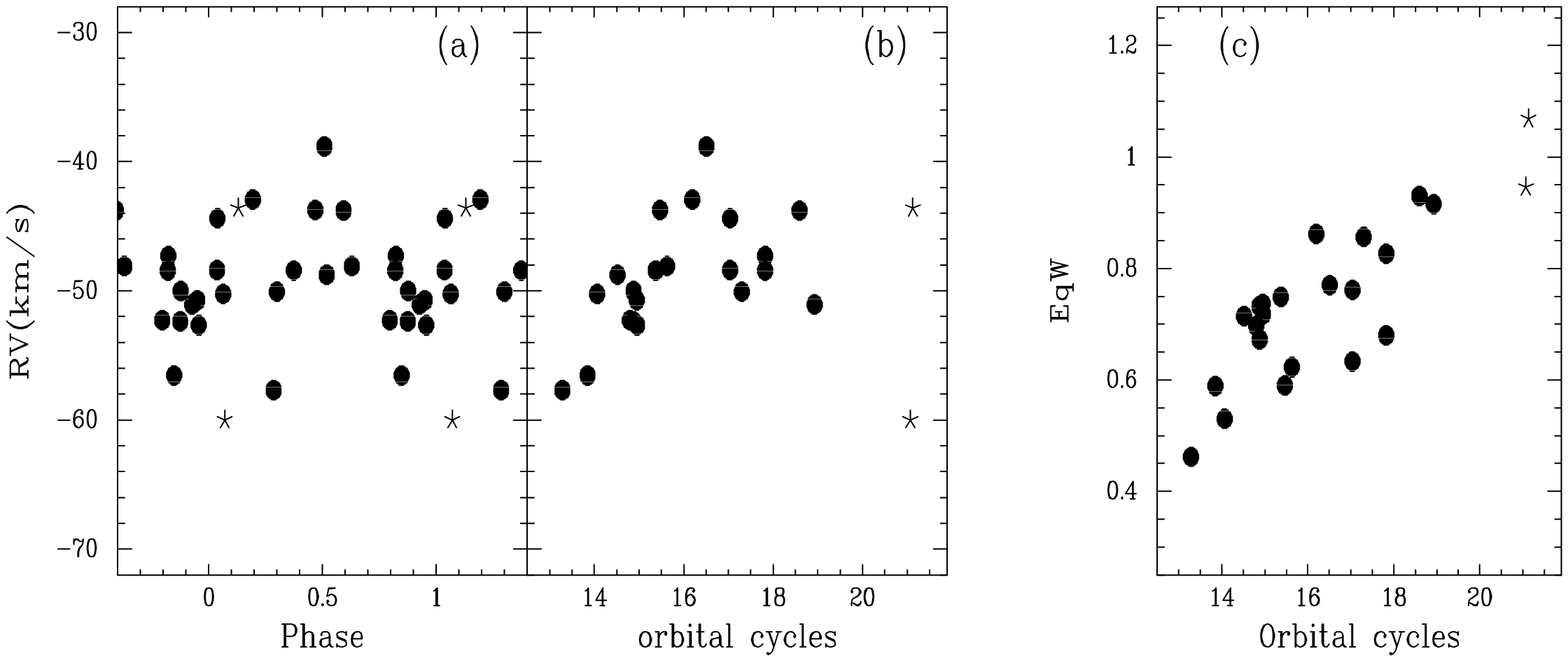}
    \includegraphics[width=5cm,angle=90]{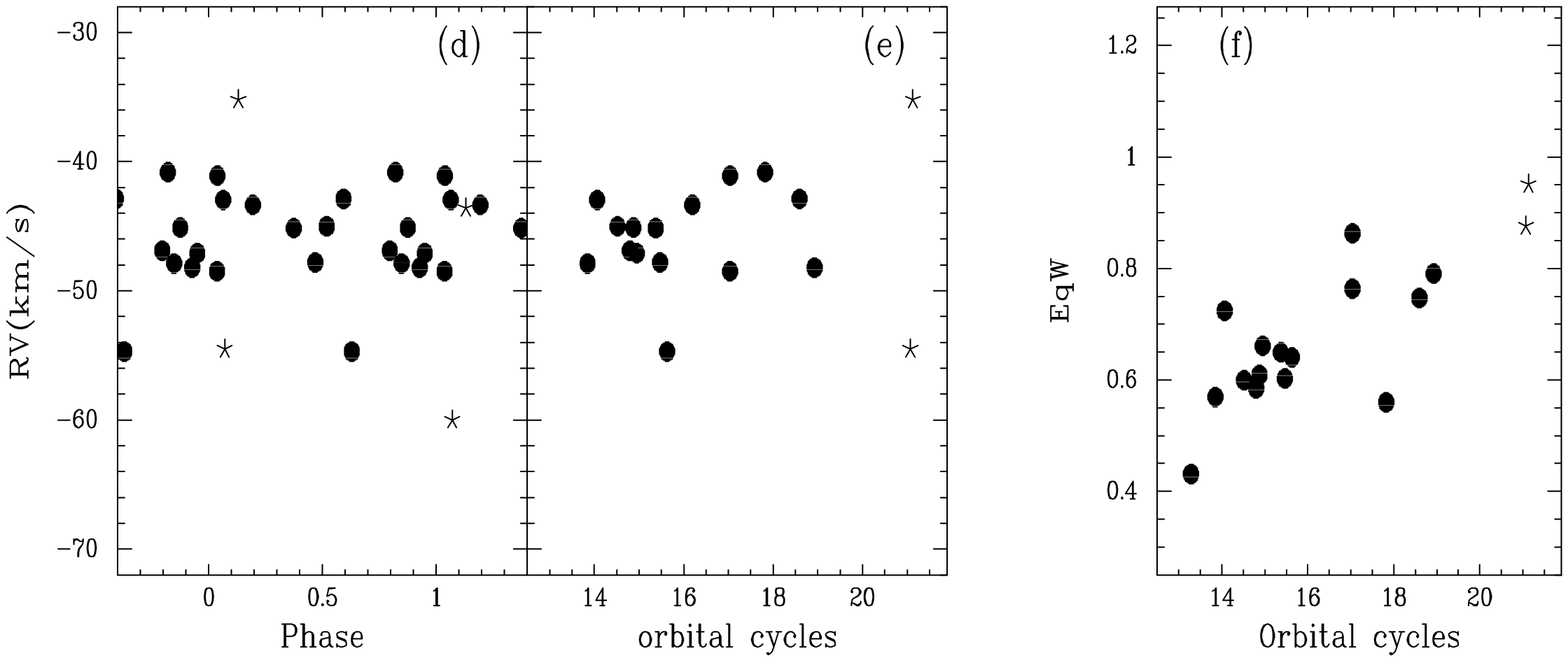}
   \caption {Radial velocities and equivalent widths of the central absorptions of H$\alpha$ 
({\it panels (a),(b) and (c)})
and H$\beta$ ({\it panels (d), (e) and (f)}) in function of the orbital phases and the 
orbital cycles, respectively. Asterisks correspond to the Feros spectra.
}
  \label{fig.5}
    \end{figure*}

The spectra present essentially emission lines of \ion{H}{i}, \ion{He}{i} and 
[\ion{O}{i}] at 6300 \AA.\,
In the observed members of the Balmer series, the broad emission is cut by a central 
absorption with a red peak always stronger than the blue one. 
Fig.~\ref{fig.4}, left and central panels, show H$\alpha$ profiles for 
several epochs. At quiescence the central absorption remains present although with 
variable intensity over the
whole orbital cycle,  whereas it disappears during the eruption, and it is weakly
visible again in October 2006. The radial velocity of the central absorption of H$\alpha$ 
varies between -60 and -40 $\rm km\,s^{-1}$. As it is shown in Fig.~\ref{fig.5}(a)
this variation does not follow the motion of any stellar component of the system but 
Fig.~\ref{fig.5}(b) and (c)
show changes of both, the radial velocities and the equivalent widths with the different 
cycles of the binary, in the sense that the mean velocity 
was increasing until cycle 16 (May 2003) and then decreasing whereas the equivalent widths 
were monotonically increasing. \\
The profile of H$\beta$  also presents a double peak structure but the radial velocity of 
the central absorption is rather constant, without a clear variation with
orbital phase or orbital cycle (see Fig.~\ref{fig.5}(d), (e) and (f)).

Significant changes in the radial velocity of the central absorption in both lines,
H$\alpha$ and H$\beta$ were observed in the  higher resolution Feros spectra 
between May and June 2008 
(asterisks in Fig.~\ref{fig.5}). An explanation of this behaviour correlating the 
radial velocities with the flickering activity and the accretion processes in RS Oph 
is in progress for a next paper.

Broad emission components were detected in the emission line \ion{He}{i}\, 
$\lambda$5875 during April 2004 and September 2005 (see Fig.~\ref{fig.4}, right panel) 
at velocities of the order of $\sim 100\, \rm km\,s^{-1}$.
The intensity of the \ion{He}{i} emission lines and the presence of these broad components 
change quickly with a time scale of one or two days.

\begin{figure}
   \centering
   \includegraphics[width=7cm]{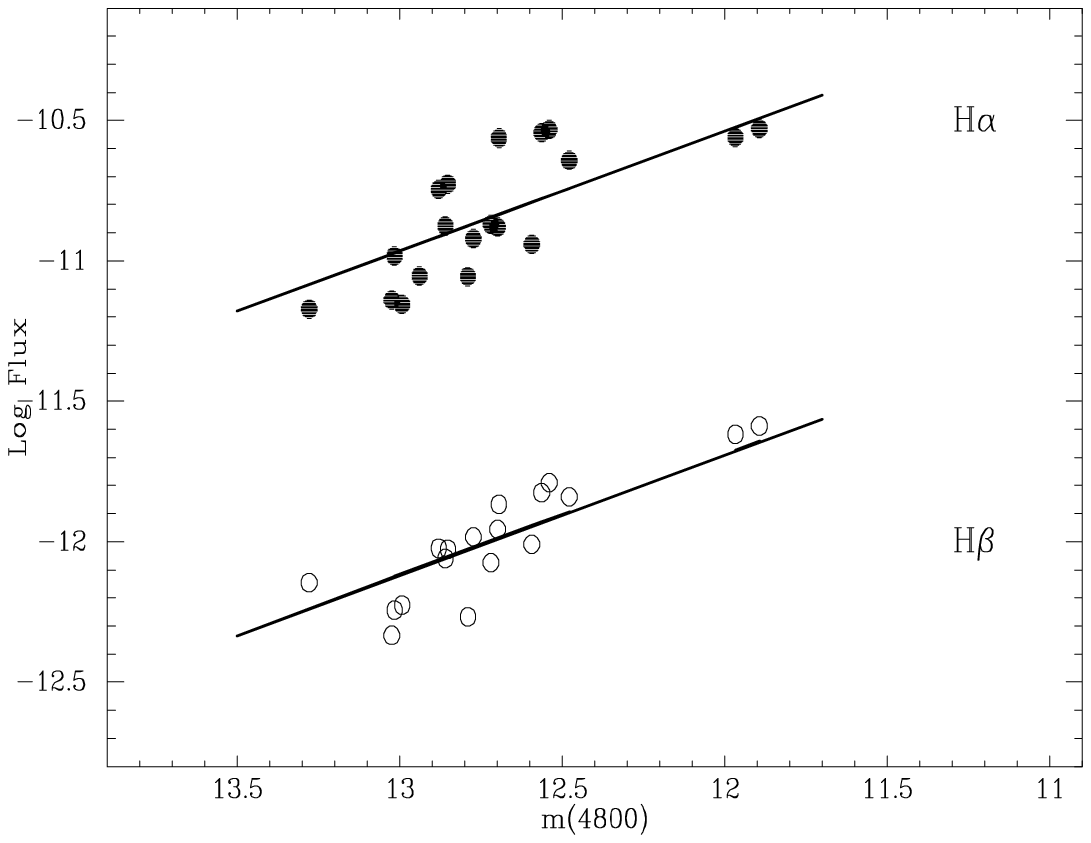}
      \caption{Correlation of the H$\alpha$ and H$\beta$ emission line fluxes
with the magnitudes at 4800 \AA\ measured from our spectra. 
              }
         \label{fig.6}
   \end{figure}

Fig.~\ref{fig.6} presents the H$\alpha$ fluxes against the  $m_{4800}$ magnitudes calculated 
from our spectra. The figure shows that a clear correlation between both data exists,
confirming that the emission line variability  is correlated with the hot component 
activity (Anupama \&  Miko{\l}ajewska, 1999).\\

%                                                One column figures
%_______________________________________________FIG.5

%______________________________________________________FIG.6

%______________________________________________________________

\subsection{Presence of the \ion{Li}{i}\, $\lambda$6707  absorption line}

One unsolved problem in astrophysics refers to the origin of lithium and 
others light elements in the universe. How and when lithium is depleted
and an understanding of the processes controlling lithium production 
are still not completely solved. The synthesis of $^{7}$Li in explosive 
hydrogen burning, in particular in accreting (CO or ONe) white dwarfs 
exploding as classical novae has been studied by several authors 
(see review Jordi (2005) and references therein) and novae are predicted 
as important sources of Galactic lithium.

However the observation of $^{7}$Li in the ejecta of a nova outburst has 
been rather elusive. 
\ion{Li}{i}\, $\lambda6708$ line  has been detected in the recurrent 
symbiotic novae T CrB (Shabhaz et al. 1999) and RS Oph (Wallerstein et al. 2006). 
Wallerstein et al.(2008) have recently derived Li abundances in these stars, $\log N(\ion{Li})=1.2$ for RS  Oph and 0.8 for T CrB, which are close to being solar. These Li abundances are however significantly higher than those determined for most single K and M giants. 
Such high Li abundances are common in late-type secondaries in neutron 
and black hole binaries (e.g. Martin et al. 1994) but extremely rare in the 
symbiotic giants. In fact, the Li enhancement 
is thus far detected only in the  symbiotic Mira V407 Cyg, where it 
can probably be explained as a consequence of hot bottom burning, which 
occurs in stars with initial masses in the range 4-6 $\rm M_{\sun}$ 
(Tatarnikova et al. 2003). Such an explanation is, however, unlikely in the 
case of low mass, $\la 1 \rm M_{\sun}$, nonpulsating giants in RS Oph and T CrB.

\begin{figure}
   \centering
   \includegraphics[width=6cm,angle=90]{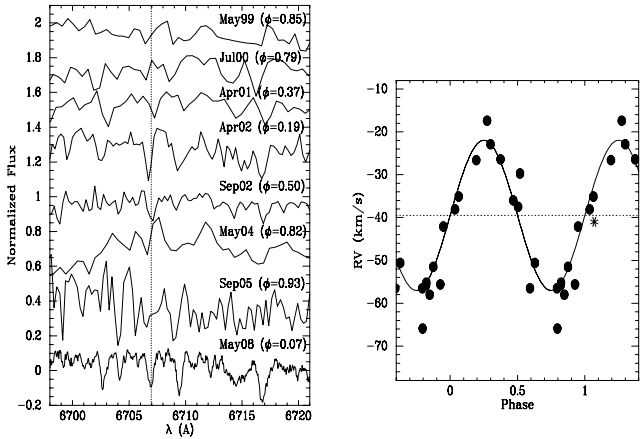}
      \caption{({\it left}) Presence of the \ion{Li}{i}\,$\lambda$6707 
absorption 
line in RS Oph during quiescence. The dotted line indicates the position of 
the line corrected by the baricentric velocity of the binary.
({\it right}) The radial velocities of \ion{Li}{i} following the M-giant 
motion.
Filled circles, Casleo data and asterisk, Feros data. 
Solid line gives the circular solution for the M-giant, as {\bf in} Fig.~\ref{fig.1}.
              }
         \label{fig.7}
   \end{figure}

Our spectra of RS Oph reveal also the presence of the \ion{Li}{i} line, and 
the radial velocities 
of this line fit very well the M-giant radial velocity curve (see right panel 
Fig.~\ref{fig.7}). 
The identification of the weak and narrow absorption line at 6706-6708 \AA\, with 
\ion{Li}{i}\,$\lambda$6708 is unambiguous since no other neutral element  
following the giant motion has a strong transition at that wavelength. 
In addtition, the presence of this line is clearly seen in the higher resolution 
Feros spectra. 
The feature was observed in all our spectra, except in April 2006 where all the 
absorption lines were 
overwhelmed by very strong blue continuum from the nova outburst.

This result indicates that, regardless of the mechanism  involved 
to generate lithium, it should operate within the cool giant atmosphere. 
The observed Li would be the initial Li (which is observed in non convective stars) 
if  some process operating in such binary systems as RS Oph and T CrB, e.g. 
delayed onset of convection or a lack of differential rotation due to tidal locking,  
can inhibit lithium depletion (Shabhaz et al. 1999). 

 Wallerstein et al. (2008) offered another
possibility for the presence of Li. They suggested that it is freshly
created in the interior and convected to the surface. They also noted that some G and K-type
giants have very high atmospheric Li abundances. In the case of these Li-rich giants,
Charbonnel \& Balachandran (2000) identified two distinct episodes of Li production occuring in advanced evolutionary phases depending upon the mass of the star. Low-mass RGB stars which later undergo the helium flash, produce Li at the start of the red giant luminosity bump phase whereas in intermediate-mass stars  the Li-rich phase occurs when the convective envelope deepens at the base of the AGB.
The position of the cool giant of T CrB in the HR diagram is consistent with a low-mass, $\sim 1\, \rm M_{\odot}$, star at the top of RGB whereas the location  of RS Oph is consistent with such a giant only if its metallicity is significantly subsolar (Miko{\l}ajewska 2008), and in both cases the stars are located far from the RGB bump.
On the other hand, Wallerstein et al. (2008) found a normal, nearly solar, ratio of metals to hydrogen in both RS Oph and T CrB, and they concluded that the giant component has not yet lost sufficient mass to make its atmosphere deficient in hydrogen. It seems thus unlikely that the initial mass of the M giant was significantly greater than its current estimate of $0.68--0.8\, \rm M_{\odot}$.

An alternative explanation is the polution of the giant by the nova ejecta. 
This explanation has, 
however,  some weak points. In particular, the red giant should very efficiently 
accrete the Li-rich 
material ejected at very high velocity (a few 1000 $\rm km\,s^{-1}$!) to be significantly poluted. 
Moreover there is also serious controversy about the production of Li during such an 
explosion.

%____________________________________________________________FIG.7
%______________________________________________________________

\section {RS Oph in 2006}

High-dispersion spectra of RS Oph were taken at CASLEO with the same 
instrumental configuration on 2006 April 7--8 and October 12, that is, 54--55 
and 240 days following the explosion, respectively.
 The spectra of April show broad emission lines of hydrogen recombination 
lines, together with \ion{He}{i}, \ion{He}{ii}, \ion{Fe}{ii}, \ion{N}{iii}, 
[\ion{O}{i}]\, $\lambda$6300, [\ion{O}{iii}]\, $\lambda$5007, 
[\ion{N}{ii}]\, $\lambda$5754 and the Raman band at $\lambda$6825 \AA\ . 
The cF absorption system is not observed in April 2006.
Very narrow emission components are seen on the top of the broad 
emission components.

\begin{figure}
   \centering
   \includegraphics[width=5.5cm,angle=90]{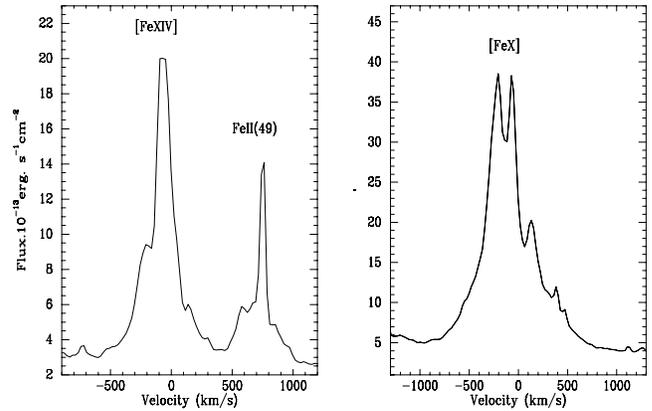}
      \caption{Observed coronal lines in RS Oph on April 2006.
              }
         \label{fig.8}
   \end{figure}

Strong coronal emission lines such as [\ion{Fe}{xiv}]\, $\lambda$5305, 
[\ion{A}{x}]\, $\lambda$5535 and [\ion{Fe}{x}]\, $\lambda$6375 
(Fig.~\ref{fig.8}) are also present, showing the same structure in the profiles.
Our measures of the integrated fluxes, the radial velocity of the emission line 
components and the full width at zero intensity (FWZI) of several stronger lines 
are shown in Table~\ref{apr06-em}.

%____________________________________________________________FIG.8

%______________________________________________________________

%____________________________________________________________FIG.9

 \begin{figure}
   \centering
   \includegraphics[width=8cm]{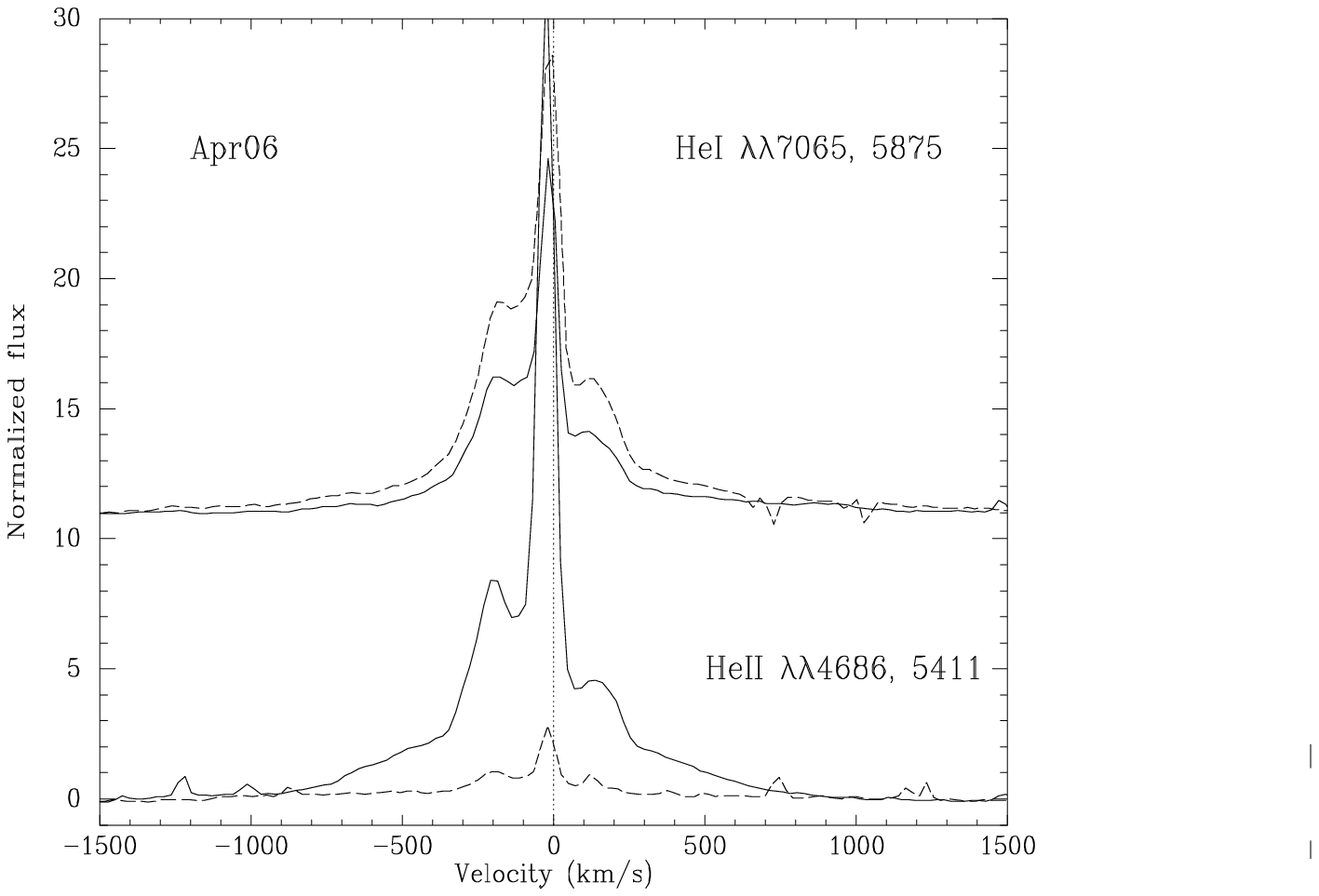}
      \caption{The He\,I and He\,II profiles showing a broad pedestal with the 
        strong narrow component at -15 and -23 $\rm km\,s^{-1}$ respectively. 
       Two separate emission components of hiher radial velocity at -200 and +150
       $kms^{-1}$ are observed being the blue component stronger than the red one.      
              }
         \label{fig.9}
   \end{figure}
%______________________________________________________________

%___________________________________________________________FIG.10

 \begin{figure}
   \centering
   \includegraphics[width=5cm,angle=90]{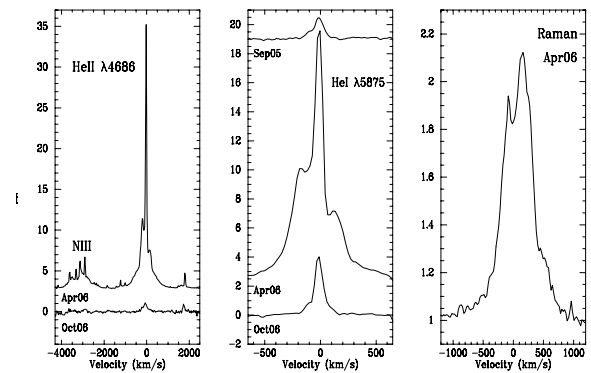}
      \caption{({\it left panel}): N\,III emission lines are observed on
April 2006. The He\,II$\lambda$4686 is still detected on 2006 
October 12. ({\it middle panel}): intensity and profile variations 
of He\,I$\lambda$5875 before and after explosion. ({\it right panel}):
 the scattering Raman band at 6825 A, which does not
appear at all during the quiescent phase, is present as a strong
emission profile in April 2006. This feature is not observed in 
October 2006.  
              }
         \label{fig.10}
   \end{figure}
%______________________________________________________________

Most of the emission lines present a strong and narrow component with a 
radial velocity between -14 and -25 $\rm km\,s^{-1}$ and two other components 
extended to the blue and the red side.
In the cases of the [\ion{N}{ii}] and [\ion{O}{iii}] emission lines and 
the coronal lines, the radial velocity of the narrow component is more 
negative (-40 and -50 $\rm km\,s^{-1}$).

We can see in Fig.~\ref{fig.9} the \ion{He}{i}\, $\lambda\lambda$5875,7065  and 
\ion{He}{ii}\, $\lambda$4686, 5411 profiles showing a broad pedestal with 
the strong narrow component at -15 and -23 $\rm km\,s^{-1}$ respectively. Two separate 
emission components 
of higher radial velocity are observed at $\sim$-200 and +150 $\rm km\,s^{-1}$ 
with the blue component stronger than the red one.
An explanation based on a bipolar gas outflow may be applicable to the presence of these
components with expanding and receding radial velocities.\\

On October 2006, the RS Oph spectrum is slowly restoring its quiescent characteristics. 
The cF-absorption system, absent in April 2006,  has reappeared.
The intensity of the emission lines decreases, the coronal lines are not observed 
and the blue and red components of the \ion{He}{i} lines have disappeared. H$\alpha$ has 
recovered its pre-outburst profile, in particular the emission wings are narrower than in 
April 2006 and the emission is cut by an incipient central absorption (see Fig.~\ref{fig.4}). 
A very weak \ion{He}{ii}\, $\lambda$4686 line is still observed, although its flux has 
decreased a factor of $\sim$90 with respect to that of April 2006. This profile 
preserves the strong emission component at $-46\, \rm km\,s^{-1}$ and two expanding 
components at $-214$ and +192 $\rm km\,s^{-1}$, with the red component being stronger 
than the blue one, that is, in opposite sense of that of April (Fig.~\ref{fig.10}, 
left panel).

% ------------------------------------------------TABLE 4

%__________________________________________________ One column table
\begin{table}
\caption{Emission lines of RS Oph on April 2006}
\tabcolsep1pt
\label{apr06-em}
\begin{tabular}{lccccccc} 
\hline
Species & $\lambda$ & Flux ($\rm 10^{-12}$) &\multicolumn{3}{c}{RV (km$\,s^{-1}$)}& FWZI     \cr
        & [\AA]     & (erg\,s$^{-1}\, cm^{-2}$)& narrow   &blue  & red &  (km\,s$^{-1}$) \cr 
\hline  
 H\,I &        &                                             &      &  &        \cr 
H$\beta$ & 4861    & 59.1 &   -21  & -38 &         & 2500    \cr  
H$\alpha$& 6562    & 460  &   -47  &     &         & 3840        \cr
P23      & 8345    & 0.6 &    -19  &     &         &     \cr
P22      & 8359    & 1.0 &    -20  &     &         &     \cr
P21      & 8374    & 1.0 &    -23  &     &         &   \cr 
P20      & 8392    & 1.3 &    -20  &      &       &     \cr
P19      & 8413    & 2.2 &    -16  &     &         &  \cr
\hline
\ion{He}{i} &      &      &         &      &     &       \cr 
         & 4471   & 2.6 &    -19  & -205 &  81  &  1100\cr         
& 5875   & 25.4&    -14  & -194 & 155  &  1950   \cr
         & 6678   & 9.4 &    -15  & -208 &      &  740   \cr
         & 7065   & 28.1 &   -17  & -196 & 156  &  1400  \cr
\hline
\ion{He}{ii} &     &      &        &      &      &        \cr 
         & 4686   & 26.7 &   -25  & -209 & 164  &  1680  \cr
         & 5411   &  3.4 &   -21  & -208 & 122  &  1050     \cr
\hline
  \ion{Fe}{ii}    &     &      &        &      &      &     \cr 
  M27    & 4233   & 1.9  &  -20  & -186  &      &  800    \cr
  M38    & 4583   & 1.7  &  -20  & -272: & 77w  &  650  \cr
  M42    & 5169   & 2.3  & -20   & -121  & 123w &  580  \cr
  M49    & 5234   & 1.4  & -22   & -124  & 134  &  720  \cr
  M49    & 5316   & 2.4  & -18   & -195  & 88w  &  580  \cr
  M74    & 6247   & 0.8  & -22   & -141  & 47   &  514  \cr
  M73    & 7711   & 1.7  & -24   & -122  & 181  &  717  \cr
\hline
Raman band & 6825 & 5.1  &       &       &      & 1470  \cr
\hline
 Forbidden lines &   &      &       &    &              \cr
[\ion{O}{i}]     & 6300  & 0.6 & -15  &  -94  & 90 & 350 \cr
[\ion{N}{ii}]    & 5754  & 4.6 & -42  &       & 81 & 925 \cr
[\ion{O}{iii}]   & 5007  & 1.7 & -40  &  -104 &   &  320 \cr
\hline
Coronal lines &     &     &        &    &                   \cr
[\ion{Fe}{xiv}] & 5303 & 7.0 & -52   & -220 & 163w & 900     \cr
[\ion{A}{x}] & 5535($\star$) & 7.5 & -13 & -282 & 100w &1930  \cr
[\ion{Fe}{x}]   & 6375 & 35.1& -40   & -188 & 166 & 1500       \cr
\hline
\end{tabular}
 \begin{list}{}{}   
\item w: very weak line
\item ($\star$): blend with \ion{Fe}{ii}$\lambda$5534.9

\end{list} 
\end{table}
%\end{tabular}

%---------------------------------------------------------------

\section{Conclusions}

 \begin{enumerate}
     
\item We have re-determined the spectroscopic orbits of RS Oph based on the radial 
velocity curve of the M-type absorption lines at wavelengths longer than 6000 \AA.\, 
As the \ion{H}{i} broad emission wings seem to follow the hot component motion, 
combined orbital solutions for 
both components were also determined with a period of 453.6 days, very similar to 
those given by FJHS00.
We conclude, as DK94 do, that the cF-type absorption lines are not associated with 
either binary component, and are most likely formed in the material streaming 
towards the hot component.

Assuming a massive white dwarf as the hot component of the system 
($M_{\rm h}= 1.2 - 1.4\,\rm M_{\odot}$) the red giant mass results in
$M_{\rm g}= 0.68 - 0.80\,\rm M_{\odot}$  and the orbit inclination, 
$ i = 49^{\circ} - 52^{\circ}$.

\item During the quiescent period of our observations the spectra of RS Oph show 
variability in the fluxes of the emission lines and in the continuum. A correlation 
of the \ion{H}{i} and \ion{He}{i} emission line fluxes with the monochromatic 
magnitudes at 4800 \AA\ was obtained, indicating 
that the hot component activity is responsible of those flux variations. 

\item  Our spectra of RS Oph reveal the presence of the \ion{Li}{i}\, 6708 line, 
and the radial velocities of this line fit very well the M-giant radial velocity 
curve. This result indicates that, whatever the mechanism involved to generate lithium
would be, it should operate within the cool giant atmosphere.
 An alternative explanation is the polution of the giant by the nova ejecta 
but the red giant should very efficiently accrete in this case the Li-rich 
material ejected at very high velocity to be significantly poluted. 

\item We present the characteristics of the spectra around 54--55 and 240 days after 
the outburst of February 2006. On April 2006 the most of the emission lines present 
a broad pedestal with a strong and narrow component at about -20 $\rm km\,s^{-1}$ and two 
other extended emission components at -200 and +150 $\rm km\,s^{-1}$.  These components 
could originate in a bipolar gas outflow supporting 
the model of a bipolar shock-heated shell expanding through the cool component wind and 
perpendicular to the orbital plane of the binary.

\item Our observations indicate that the cF system was disrupted during the outburst. 
The cF absorption lines were not observed in our spectra in April 2006 and 
Zamanov et al. (2006) reported the absence of flickering in June 2006, then the 
accretion was not yet resumed immediately after the outburst.
Both, the cF system and the flickering (Worters et al. 2007) were observed again 
about 240 days after the outburst which is consistent with the resumption of accretion 
in RS Oph.

  \end{enumerate}

\begin{acknowledgements}
We are deeply indebted to Drs. N. Morrell and R. Barb\'a for obtaining the Feros spectrograms 
at ESO, La Silla (Chile).  We also thank the anonymous referee for very valuable comments.
This research was partly supported by Polish research grants No. 1P03D01727, and N203 395534.
The CCD and data acquisition system at CASLEO has been partly finance by R.M. Rich through U.S. 
NSF grant AST-90-15827.

\end{acknowledgements}

%_____________________________________________longtable-Table1
\longtab{1}{
\begin{longtable}{lccccccc}
\caption{\label{rv} Radial velocities of the M-giant component, H$\alpha$-wings and cF-absortions  in RS Oph}\\
\hline\hline
JD & Phase ($\dag$) & \multicolumn{2}{c}{M-giant} &  \multicolumn{2}{c}{H$\alpha$-wings} & cF-abs & Source\\
2400000+&  & RV(km\,s$^{-1}$) & $(O-C)$ & RV(km\,s$^{-1}$)&$(O-C)$ & RV(km\,s$^{-1})$ &  \\
\hline
\endfirsthead
\caption{continued.}\\
\hline\hline
JD & Phase ($\dag$) & \multicolumn{2}{c}{M-giant} &  \multicolumn{2}{c}{H$\alpha$-wings} & cF-abs & Source \\
2400000+&  & RV(km\,s$^{-1}$) & $(O-C)$  & RV(km\,s$^{-1}$)&$(O-C)$ & RV(km\,s$^{-1}$)&  \\
\hline
\endhead
\hline
\endfoot

45035.038 & -0.019 & -41.0 & 0.0 &   &   &   & DK94 \\
45063.017 & 0.043 & -32.0 & 2.5 &   &   &   & DK94 \\
45063.963 & 0.045 & -35.0 & -0.7 &   &   &   & DK94 \\
45097.923 & 0.120 & -27.4 & -0.1 &   &   &   & DK94 \\
45426.977 & 0.845 & -41.1 & 11.9 &   &   &   & DK94 \\
45426.993 & 0.845 & -41.6 & 11.4 &   &   &   & DK94 \\
45513.743 & 0.037 & -36.6 & -1.5 &   &   &   & DK94 \\
46538.993 & 0.297 & -29.0 & -6.8 &   &   &   & DK94 \\
46540.952 & 0.301 & -32.0 & -9.6 &   &   &   & DK94 \\
46602.834 & 0.438 & -40.7 & -8.9 &   &   &   & DK94 \\
46605.743 & 0.444 & -39.3 & -6.8 &   &   &   & DK94 \\
46608.764 & 0.451 & -39.4 & -6.3 &   &   &   & DK94 \\
46635.720 & 0.510 & -37.6 & 1.8 &   &   &   & DK94 \\
46691.608 & 0.633 & -54.1 & -3.0 &   &   &   & DK94 \\
46867.010 & 0.020 & -38.0 & -1.1 &   &   &   & DK94 \\
46891.957 & 0.075 & -34.9 & -3.7 &   &   &   & DK94 \\
46895.935 & 0.084 & -33.9 & -3.5 &   &   &   & DK94 \\
46918.913 & 0.134 & -32.2 & -6.0 &   &   &   & DK94 \\
46929.994 & 0.159 & -27.7 & -3.2 &   &   &   & DK94 \\
46953.833 & 0.211 & -23.5 & -1.3 &   &   &   & DK94 \\
46979.733 & 0.268 & -25.8 & -4.1 &   &   &   & DK94 \\
47312.835 & 0.003 & -39.3 & -0.6 &   &   &   & DK94 \\
47313.822 & 0.005 & -38.6 & -0.1 &   &   &   & DK94 \\
47486.561 & 0.386 & -38.2 & -11.0 &   &   &   & DK94 \\
47719.764 & 0.900 & -59.2 & -10.2 &   &   &   & DK94 \\
47777.656 & 0.028 & -36.0 & 0.1 &   &   &   & DK94 \\
47787.683 & 0.050 & -37.4 & -3.6 &   &   &   & DK94 \\
47811.576 & 0.102 & -30.5 & -1.8 &   &   &   & DK94 \\
47967.019 & 0.445 & -29.5 & 3.1 &   &   &   & DK94 \\
47992.928 & 0.502 & -43.9 & -5.4 &   &   &   & DK94 \\
48016.933 & 0.555 & -44.5 & -0.4 &   &   &   & DK94 \\
48048.826 & 0.625 & -54.3 & -3.8 &   &   &   & DK94 \\
48349.954 & 0.289 & -22.8 & -0.8 &   &   &   & DK94 \\
48368.857 & 0.331 & -21.2 & 2.4 &   &   &   & DK94 \\
48400.832 & 0.401 & -20.1 & 8.4 &   &   &   & DK94 \\
48411.892 & 0.426 & -30.3 & 0.4 &   &   &   & DK94 \\
48429.805 & 0.465 & -38.5 & -3.9 &   &   &   & DK94 \\
48695.025 & 0.050 & -39.5 & -5.8 &   &   &   & DK94 \\
48723.000 & 0.112 & -30.6 & -2.7 &   &   &   & DK94 \\
48754.922 & 0.182 & -28.9 & -5.6 &   &   &   & DK94 \\
48782.831 & 0.244 & -34.3 & -12.7 &   &   &   & DK94 \\
48816.771 & 0.318 & -29.5 & -6.5 &   &   &   & DK94 \\
48872.718 & 0.442 & -34.7 & -2.5 &   &   &   & DK94 \\
48909.584 & 0.523 & -44.4 & -3.6 &   &   &   & DK94 \\
49106.948 & 0.958 & -42.8 & 0.7 &   &   &   & DK94 \\
49136.907 & 0.024 & -36.6 & -0.2 &   &   &   & DK94 \\
49174.784 & 0.108 & -30.4 & -2.1 &   &   &   & DK94 \\
49998.639 & 0.924 & -49.1 & -2.3 &   &   &   & FJHS00\\ 
50162.905 & 0.286 & -18.0 & 4.0 &   &   &   & FJHS00\\ 
50254.723 & 0.488 & -33.5 & 3.6 &   &   &   & FJHS00\\ 
50320.643 & 0.634 & -48.5 & 2.6 &   &   &   & FJHS00\\ 
50387.593 & 0.781 & -61.9 & -6.4 &   &   &   & FJHS00\\ 
50568.854 & 0.181 & -27.5 & -4.2 &   &   &   & FJHS00\\ 
50627.799 & 0.311 & -19.3 & 3.4 &   &   &   & FJHS00\\ 
50751.574 & 0.584 & -49.6 & -2.7 &   &   &   & FJHS00\\ 
50933.851 & 0.986 & -42.1 & -1.5 &   &   &   & FJHS00\\ 
50981.817 & 0.091 & -27.5 & 2.2 &   &   &   & FJHS00\\ 
50983.747 & 0.096 & -26.8 & 2.5 &   &   &   & FJHS00\\ 
51051.677 & 0.245 & -19.7 & 1.9 &   &   &   & FJHS00\\ 
51064.570 & 0.274 &   &   & -40.7 & 8.0 &   & CASLEO \\
51066.532 & 0.278 & -18.4$\pm$8.3 (4) & 3.4 &   &   & -23.3$\pm$1.2 (3) & CASLEO \\
51069.467 & 0.285 & -15.3$\pm$5.4 (4) & 6.6 &   &   & -28.7$\pm$3.5 (6) & CASLEO \\
51069.537 & 0.285 & -13.9$\pm$7.5 (4) & 8.0 &   &   & -24.4$\pm$4.9 (5) & CASLEO \\
51069.571 & 0.285 & -18.6$\pm$5.7 (5) & 3.3 &   &   & -26.7$\pm$3.6 (5) & CASLEO\\ 
51106.592 & 0.367 & -27.4 & -1.7 &   &   &   & FJHS00\\ 
51294.889 & 0.782 & -63.5 & -8.0 &   &   &   & FJHS00\\ 
51324.666 & 0.847 & -55.3$\pm$5.7 (24) & -2.5 & -32.8 & -2.5 & -30.2$\pm$3.7 (4) & CASLEO \\
51324.678 & 0.847 & -53.1$\pm$2.1 (21) & -0.3 & -34.2 & -3.9 & -30.1$\pm$2.5 (5) & CASLEO \\
51324.690 & 0.847 & -54.6$\pm$2.0 (20) & -1.8 & -32.1 & 1.8 & -24.9$\pm$3.3 (3) & CASLEO \\
51363.664 & 0.933 & -45.0 & 0.9 &   &   &   & FJHS00\\ 
51422.584 & 0.063 & -33.6$\pm$1.4 (9) & -1.2 & -49.6 & -7.2 & -34.4$\pm$1.2 (7) & CASLEO \\
51422.599 & 0.063 & -29.5$\pm$2.1 (6) & 2.9 & -46.2 & -3.8 & -36.2$\pm$2.5 (7) & CASLEO \\
51628.861 & 0.518 & -37.5$\pm$1.7 (31) & 2.7 & -39.7 & -2.0 & -42.4$\pm$1.8 (7) & CASLEO \\
51628.870 & 0.518 & -38.4$\pm$1.7 (28) & 1.8 & -38.3 & -0.6 & -44.0$\pm$2.0 (6) & CASLEO \\
51629.850 & 0.520 & -39.6$\pm$2.2 (22) & 0.9 & -38.9 & -1.3 & -40.6$\pm$3.1 (7) & CASLEO \\
51629.859 & 0.520 & -39.8$\pm$1.5 (28) & 0.7 & -38.9 & -1.3 & -43.6$\pm$2.4 (7) & CASLEO \\
51629.868 & 0.520 & -39.6$\pm$1.9 (26) & 0.9 & -38.9 & -1.3 & -43.9$\pm$2.7 (6) & CASLEO \\
51629.876 & 0.520 & -40.7$\pm$2.0 (26) & -0.2 & -36.0 & 1.6 & -43.5$\pm$3.3 (6) & CASLEO \\
51630.809 & 0.522 & -37.6$\pm$1.8 (27) & 3.1 & -33.1 & 4.4 & -47.2$\pm$0.5 (4) & CASLEO \\
51630.832 & 0.522 & -39.0$\pm$2.2 (22) & 1.7 & -31.5 & 6.0 & -46.7$\pm$3.4 (5) & CASLEO \\
51754.660 & 0.795 & -59.4$\pm$3.8 (6) & -4.3 & -27.4 & 1.5 & -29.4$\pm$1.6 (7) & CASLEO \\
51754.669 & 0.795 & -58.7$\pm$3.8 (6) & -3.6 & -29.1 & -0.2 & -30.1$\pm$2. (6) & CASLEO \\
51754.680 & 0.795 & -57.7$\pm$4.1 (6) & -2.6 & -27.8 & 1.1 & -26.5$\pm$2.5 (7) & CASLEO \\
51790.615 & 0.874 & -52.4$\pm$2.9 (10) & -1.4 & -34.2 & -2.9 & -18.6$\pm$0.7 (7) & CASLEO \\
51790.623 & 0.874 & -53.1$\pm$2.7 (9) & -2.1 & -34.0 & -2.7 & -19.1$\pm$1.1 (6) & CASLEO \\
51790.630 & 0.874 & -50.2$\pm$3.8 (9) & 0.8 & -33.9 & -2.6 & -20.2$\pm$1.7 (7) & CASLEO \\
51791.616 & 0.877 & -48.4$\pm$1.1 (42) & 2.4 & -23.2 & 8.2 &   & CASLEO \\
51791.623 & 0.877 & -46.7$\pm$1.3 (39) & 4.2 & -23.7 & 7.7 &   & CASLEO \\
51791.630 & 0.877 & -47.4$\pm$1.3 (39) & 3.5 & -23.5 & 7.9 &   & CASLEO \\
51798.524 & 0.892 & -54.8$\pm$2.1 (7) & -5.1 & -29.0 & 3.2 & -18.6$\pm$4.0 (7) & CASLEO \\
51798.533 & 0.892 & -54.8$\pm$2.5 (6) & -5.1 & -26.9 & 5.3 & -22.1$\pm$1.4 (6) & CASLEO \\
51824.502 & 0.949 & -40.8$\pm$1.3 (10) & 3.6 & -29.1 & 6.2 & -32.7$\pm$1.1 (7) & CASLEO \\
51824.511 & 0.949 & -38.5$\pm$1.9 (9) & 5.9 & -28.9 & 6.4 & -32.7$\pm$1.6 (7) & CASLEO \\
51824.520 & 0.949 & -41.4$\pm$1.9 (10) & 3.0 & -28.5 & 6.8 & -33.7$\pm$1.3 (6) & CASLEO \\
51827.511 & 0.956 & -40.4$\pm$1.2 (39) & 3.3 & -30.0 & 5.7 &   & CASLEO \\
51827.518 & 0.956 & -39.0$\pm$1.1 (38) & 4.7 & -30.2 & 5.5 &   & CASLEO \\
51827.526 & 0.956 & -41.2$\pm$1.3 (38) & 2.5 & -30.2 & 5.5 &   & CASLEO \\
52016.843 & 0.373 & -19.8$\pm$2.4 (11) & 6.4 & -51.5 & -5.4 & -32.4$\pm$1.5 (7) & CASLEO \\
52016.853 & 0.373 & -24.8$\pm$1.8 (6) & 1.4 & -52.2 & -6.1 & -32.3$\pm$2.6 (7) & CASLEO \\
52016.861 & 0.373 & -24.3$\pm$4.6 (5) & 1.9 & -49.4 & -3.3 & -35.6$\pm$1.6 (6) & CASLEO \\
52058.859 & 0.466 & -36.6$\pm$4.3 (9) & -1.9 & -34.8 & 6.2 &   & CASLEO \\
52058.867 & 0.466 & -37.9$\pm$9.7 (6) & -3.2 & -34.3 & 6.7 &   & CASLEO \\
52058.875 & 0.466 & -34.0$\pm$2.1 (2) & 0.7 & -38.3 & 2.7 &   & CASLEO \\
52059.853 & 0.468 & -33.5$\pm$1.2 (36) & 1.4 &   &   &   & CASLEO \\
52059.861 & 0.468 & -34.0$\pm$1.1 (33) & 0.9 & -39.0 & 1.9 &   & CASLEO \\
52059.869 & 0.468 & -34.3$\pm$1.1 (32) & 0.6 & -40.1 & 0.8 &   & CASLEO \\
52132.575 & 0.628 & -60.7$\pm$1.2 (7) & -10.0 & -34.1 & -2.6 & -41.8$\pm$5.1 (6) & CASLEO \\
52132.584 & 0.628 & -61.0$\pm$2.0 (7) & -10.3 & -35.0 & -3.5 & -42.1$\pm$5.1 (6) & CASLEO \\
52388.895 & 0.193 & -26.8$\pm$1.0 (89) & -4.0 & -50.5 & -2.4 & -33.5$\pm$0.2 (2) & CASLEO \\
52388.914 & 0.193 & -24.4$\pm$0.95 (93) & -1.6 & -50.2 & -2.1 & -26.7$\pm$1.2 (2) & CASLEO \\
52529.553 & 0.504 &   &   & -28.7 & 10.0 &   & CASLEO \\
52531.543 & 0.508 & -38.7$\pm$1.3 (122) & 0.5 & -32.7 & 5.7 &   & CASLEO \\
52531.556 & 0.508 & -37.7$\pm$1.3 (123) & 1.5 & -33.7 & 4.7 &   & CASLEO \\
52770.858 & 0.036 & -34.5$\pm$1.6 (70) & 0.7 & -28.7 & 12.0 & -26.9$\pm$2.2 (3) & CASLEO \\
52770.873 & 0.036 & -37.2$\pm$1.8 (68) & -2.0 & -29.2 & 11.5 & -26.1$\pm$0.4 (3) & CASLEO \\
52771.837 & 0.038 & -36.6$\pm$3.2 (29) & -1.6 & -35.5 & 5.3 & -26.9$\pm$1.8 (6) & CASLEO \\
52771.846 & 0.038 & -37.2$\pm$3.3 (24) & -2.2 & -36.6 & 4.2 & -25.3$\pm$1.8 (6) & CASLEO \\
52772.848 & 0.040 &   &   & -31.4 & 9.6 &   & CASLEO \\
52772.860 & 0.040 &   &   & -30.7 & 10.3 &   & CASLEO \\
52887.575 & 0.292 &   &   & -38.7 & 9.8 &   & CASLEO \\
52890.529 & 0.299 & -21.6$\pm$0.9 (110) & 0.7 & -34.4 & 14.0 &   & CASLEO \\
52890.542 & 0.299 & -23.5$\pm$0.9 (121) & -1.2 & -34.9 & 13.5 &   & CASLEO \\
53126.836 & 0.820 & -54.6$\pm$1.4 (65) & -0.4 &   &   & -26.5$\pm$3.4 (5) & CASLEO \\
53126.848 & 0.820 & -54.5$\pm$1.9 (55) & -0.3 &   &   & -24.5$\pm$1.6 (6) & CASLEO \\
53127.850 & 0.822 & -56.1$\pm$1.1 (121) & -2.0 & -30.9 & -1.4 &   & CASLEO \\
53127.864 & 0.822 & -54.2$\pm$1.4 (119) & -0.1 & -32.0 & -2.5 &   & CASLEO \\
53476.884 & 0.592 & -51.6$\pm$1.7 (69) & -3.9 & -28.0 & 5.3 & -43.5$\pm$2.7 (3) & CASLEO \\
53476.897 & 0.592 & -50.4$\pm$1.5 (74) & -2.7 & -25.4 & 7.9 & -42.4$\pm$2.4 (3) & CASLEO \\
53479.879 & 0.599 & -55.4$\pm$1.3 (91) & -7.1 & -36.4 & -3.4 &   & CASLEO \\
53628.562 & 0.926 & -51.1$\pm$1.8 (32) & -4.5 &   &   & -38.7$\pm$2.7 (4) & CASLEO \\
54021.484 & 0.793 & -57.2$\pm$2.4 (58) & -2.0 & -18.7 & 10.2 & -25.3$\pm$0.4 (3) & CASLEO \\
54021.496 & 0.793 & -57.1$\pm$2.5 (66) & -1.9 & -20.0 & 8.9 & -24.2$\pm$2.2 (4) & CASLEO \\

\end{longtable}

\noindent ($\dag$)\,  We adopt the spectroscopic ephemeris:
T$_{\it conj}$= 2445043.54 + 453.6\,E
}
%____________________________________________________________________

\end{document}